\documentclass{mem}
\usepackage{natbib}\usepackage{txfonts}
\usepackage{graphicx}
\usepackage[a4paper,breaklinks,dvipdfm]{hyperref}
\idline{75}{282}
\begin{document}
\def\teff{$T\rm_{eff }$}
\def\kms{$\mathrm {km s}^{-1}$}

\title{
X-Shooter spectroscopy of brown dwarfs in the TW\,Hya association
}

   \subtitle{}

\author{
B.\,Stelzer\inst{1}
\and J.M.\,Alcal\'a\inst{2} 
\and A.\,Frasca\inst{3}
          }

  \offprints{B. Stelzer}

\institute{
INAF --
Osserv. Astronomico di Palermo, Piazza del Parlamento 1, 
90134 Palermo, Italy
\and
INAF -- 
Osserv. Astronomico di Capodimonte, Via Moiariello 16, 80131 Napoli, Italy
\and
INAF -- 
Osserv. Astrofisico di Catania, Via S. Sofia 78, 95123 Catania, Italy \\
\email{stelzer@astro.inaf.it}
}

\authorrunning{Stelzer, Alcal\'a \& Frasca}

\titlerunning{X-Shooter spectroscopy of TWA brown dwarfs}

\abstract{
We present broad-band mid-resolution X-Shooter/VLT spectra 
for four brown dwarfs of the TW\,Hya association.
Our targets comprise substellar analogs representing the different evolutionary
phases in young stellar evolution: 
In the two diskless brown dwarfs 
we determine the stellar parameters and we study 
their chromospheric emission line spectrum.
For the two accreting brown dwarfs 
we estimate the mass accretion rates.

}
\maketitle{}

\section{Introduction}

We discuss optical and near-infrared spectra
of four brown dwarf members of the TW\,Hya (TWA) association. The data have 
been obtained with the broad-band ($350-2500$\,nm) 
spectrograph X-Shooter at the VLT 
as part of the INAF consortium's Guaranteed Time observations 
(see Alcal\'a et al. 2011).  
The target list with spectral types and distances adopted from the literature
and the observing log are presented in Table~\ref{tab:obslog}.
\begin{table*}
\begin{center}
\caption{Observing log for X-Shooter spectroscopy of TWA brown dwarfs}
\label{tab:obslog}
\begin{tabular}{ccccclc}\hline
\multicolumn{1}{c}{TWA}    & \multicolumn{1}{c}{other}       & SpT & d & YSO & \multicolumn{1}{c}{Obs.date} & Slit width \\ 
\multicolumn{1}{c}{number} & \multicolumn{1}{c}{designation} &     & [pc] & Class & \multicolumn{1}{c}{[dd/mm/yyyy]} & UVB/VIS/NIR \\ \hline
26 & 2M\,J1139-3159     & M8...M9 & 42 & III & 22/03/2010 & 1.0"/0.9"/0.9" \\ 
27 & 2M\,J1207-3932     & M8      & 53 & II  & 22/03/2010; 19/04/2012 &  1.0"/0.9"/0.9"\\
28 & SSSPM\,J1102-3431  & M8.5    & 55 & II  & 22/03/2010 &  1.6"/1.5"/1.5" \\
29 & DENIS-P\,1245-4429 & M9.5    & 79 & III & 22/03/2010 &  1.6"/1.5"/1.5" \\
\hline
\end{tabular}
\end{center}
\end{table*}
X-Shooter spectroscopy enables 
a detailed characterization of young (sub)stellar objects,
including an accurate assessment of their fundamental parameters, kinematics, 
rotation, and magnetic activity. 
It provides a rich database 
of accretion diagnostics from the 
Br$\gamma$ and Pa$\beta$ lines in the near-IR to the Balmer jump in the UV 
including the full optical band with the Balmer series and He\,I lines 
and the Ca\,IRT. Finally, outflows can be traced through forbidden line 
emission.

\begin{figure*}  
\parbox{\textwidth}{
\parbox{0.5\textwidth}{
\resizebox{\hsize}{!}{\includegraphics[clip=true]{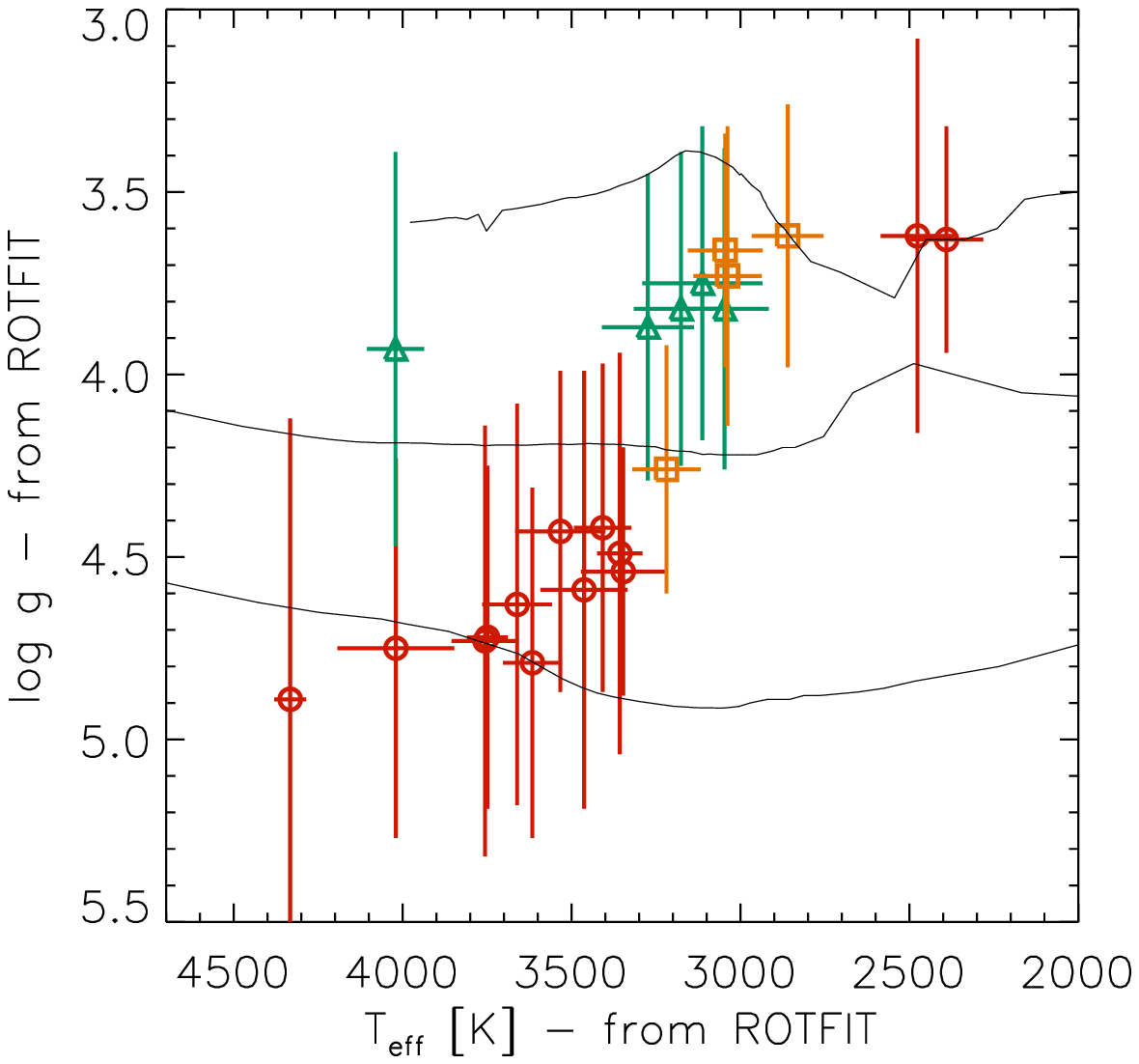}}
}
\parbox{0.5\textwidth}{
\resizebox{\hsize}{!}{\includegraphics[clip=true]{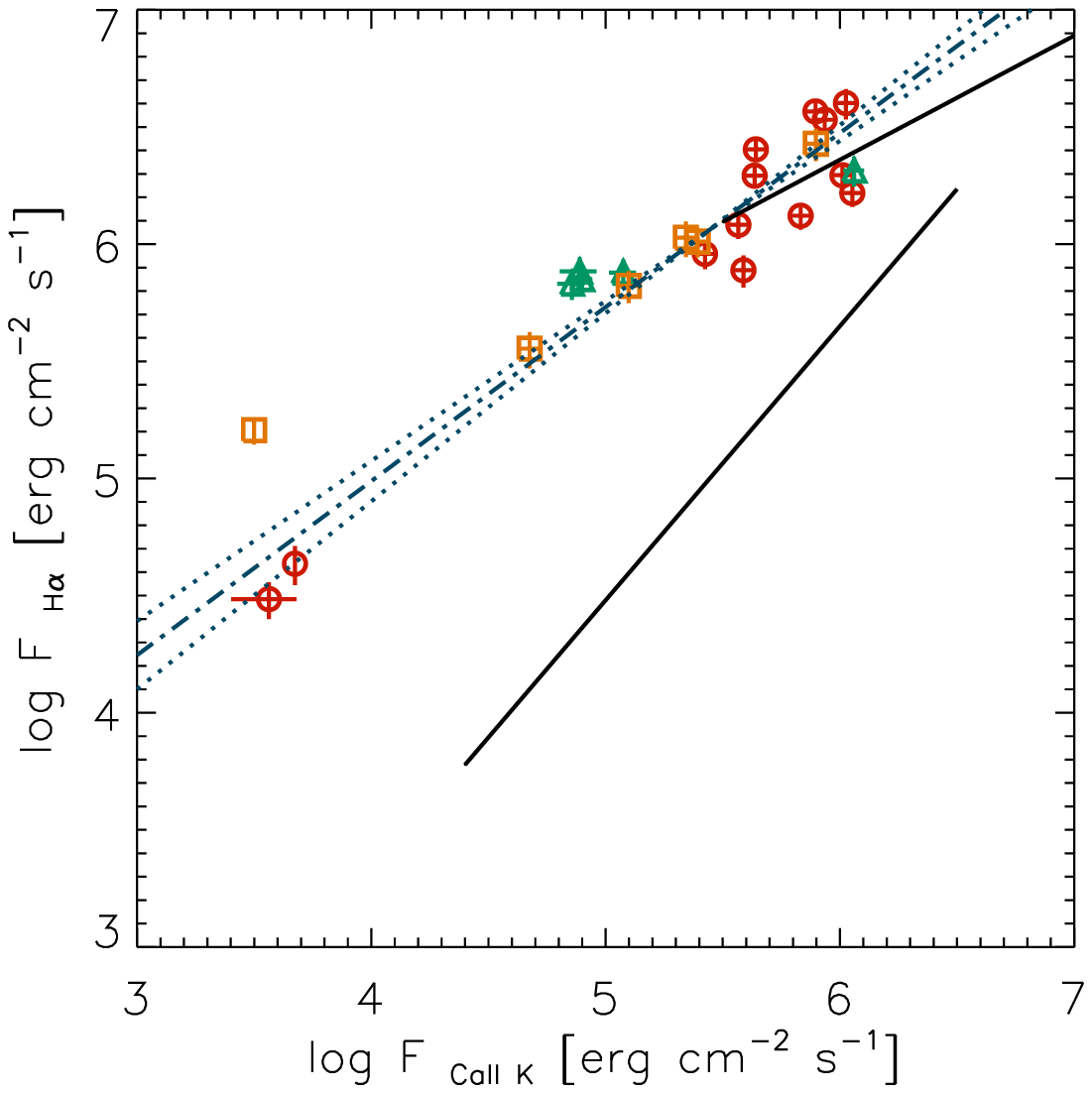}}
}
}
\caption{\footnotesize (left): Surface gravity and effective temperature measured from the X-Shooter spectrum for Class\,III sources compared to the $1$, $10$, and $100$\,Myr isochrone from \protect\cite{Baraffe98.1} and \cite{Chabrier00.2} models. (right): Flux-flux relation for chromospheric H$\alpha$ and Ca\,K emission for Class\,III sources from our X-Shooter program. Dash-dotted and dotted are the linear regression and its uncertainty. The two solid lines denote the `active' (upper) and `inactive' (lower) branches identified by MA11 for field dwarf stars. Different plotting symbols for different regions: circles - TWA, squares - Lupus, triangles - $\sigma$\,Orionis. TWA\,26 and TWA\,29 are the coolest objects with the smallest line fluxes within the TWA sample.}
\label{fig:nonacc} 
\end{figure*}

\section{Results}

Based on the presence or absence of a measurable Balmer jump we have
classified the observed TWA brown dwarfs as accretors or non-accretors, 
approximated here by the young stellar object (YSO) classes II
and III (see Table~\ref{tab:obslog}). 
The result is consistent with the previous literature where TWA\,26 was 
identified as a Class\,III source through the absence of near-IR excess
in its spectral energy distribution \citep{Morrow08.1}, while TWA\,27 and TWA\,28 display IR
excess (Riaz et al. 2006; Harvey et al. 2012). 
TWA\,27 is also known to be accreting \citep{Scholz05.4}
and to drive an outflow \citep{Whelan07.1}. 
The coolest target, TWA\,29, had not yet been investigated 
for its disk, accretion and outflow properties. 

\begin{figure*}[t!]
\resizebox{\hsize}{!}{\includegraphics[clip=true]{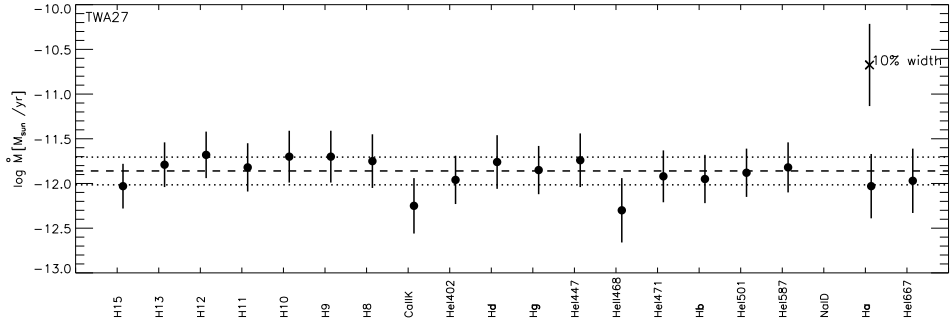}}
\caption{\footnotesize Mass accretion rate of TWA\,27 
from empirical relations between accretion luminosity 
and luminosity of emission lines. Dashed and dotted lines are mean 
($\langle \log{\dot{M}_{\rm acc,line}} \rangle = -11.9$) 
and its $1\,\sigma$ uncertainty.}
\label{fig:mdot_twa27}
\end{figure*}

\subsection{Non-accreting brown dwarfs}

For TWA\,26 and TWA\,29 and all the other Class\,III sources observed within our
X-Shooter program we have performed 
a systematic assessment of their fundamental parameters, radial velocity 
and rotational velocity using the ROTFIT routine \citep[see][]{Frasca06.0}.
In Fig.~\ref{fig:nonacc}\,(left) we show the results for surface gravity and
effective temperature compared to evolutionary models from \cite{Baraffe98.1}
and \cite{Chabrier00.2}.
Generally, we find lower gravities for the Class\,III stars in Lupus and
$\sigma$\,Orionis with respect to TWA. This is qualitatively 
consistent with the known age difference between these star forming 
regions but the values of the $\log{g}$ in TWA are higher than predicted
for $10-15$\,Myr old objects by the evolutionary models. 
The two brown dwarfs 
have lower gravity than the
early- to mid-M stars in the TWA, similar to the results of \cite{Mohanty04.2} for
a sample of brown dwarfs in the $5$\,Myr-old UpperSco region.
The temperature region around $T_{\rm eff} \approx 2500$\,K is characterized by the 
onset of dust formation and we conjecture that this aspect is not yet 
properly treated in the stellar models. 

We have measured emission lines fluxes to assess the characteristics of the
chromospheres for the Class\,III sample, including TWA\,26 and TWA\,29. 
The correlation between the surface fluxes of H$\alpha$ and Ca\,K emission 
is shown in Fig.~\ref{fig:nonacc}\,(right). 
It is different from the relation
presented by \cite{MartinezArnaiz11.0} (MA11) for the bulk of field FGKM dwarfs (lower solid 
line in Fig.~\ref{fig:nonacc}\,right). Our data are, instead, 
roughly consistent with the `active' 
branch identified by MA11 for a subsample of field M dwarfs (upper solid line) 
and we extend this upper branch to two dex lower fluxes.

%

\subsection{Accreting brown dwarfs}

For the two accretors, TWA\,27 and TWA\,28, we have measured the mass accretion rate
from empirical relations between the accretion luminosity and the emission line luminosity 
of the form $L_{\rm acc} = a + b \cdot L_{\rm line}$. Here, $L_{\rm acc}$ represents the 
UV excess emission with respect to a template spectrum of a non-accreting star of 
the same spectral type. It is obtained by adding the continuum emission of a hot slab 
of hydrogen gas to the template spectrum and fitting it to match the observed spectrum
of the accretor. The coefficients $a$ and $b$ have been determined for $39$ emission
lines from a large sample of 
accreting YSOs in Lupus observed during our program (Alcal\'a et al., in prep.). 
The mass accretion rate, $\dot{M}_{\rm acc}$, is given as 
$\dot{M}_{\rm acc} = 1.25 \cdot L_{\rm acc} R_* / G M_*$ \citep[see e.g.][]{Gullbring98.1}. 
The values obtained for TWA\,27 are shown in Fig.\ref{fig:mdot_twa27}. 
The $\dot{M}_{\rm acc}$ we derive from the H$\alpha$ $10$\,\% width using the relation
of \cite{Natta06.1} is in disagreement with the values from the line luminosities, possibly due
to outflow contributions in the H$\alpha$ wings. In an analogous way we find for TWA\,28
from the line luminosities $\langle \log{\dot{M}_{\rm acc,line}} \rangle = -12.2 \pm 0.2$
and from the H$\alpha$ $10$\,\% width $\log{\dot{M}_{\rm acc,H\alpha10\%}} = -11.3 \pm 0.4$.
We conclude that the $10$\,\% width of H$\alpha$ emission, although efficient for identifying
accretors, is not suited for quantitative measures of mass accretion. 
With these X-Shooter observations we are pushing into very low values for $\dot{M}_{\rm acc}$,
close to the limit where lines are dominated by chromospheric emission \citep{Manara13.0}.


\bibliographystyle{aa}

\end{document}